\documentclass{jpsj-suppl}
\usepackage{txfonts} 

\title{Particle Production from Geometric Transition in Expanding Universe}

\author{Sang Pyo \textsc{KIM}$^{1}$}

\inst{$^{1}$ Department of Physics, Kunsan National University, Kunsan 573-701, Korea}

\email{sangkim@kunsan.ac.kr}

\recdate{July 12, 2013}

\abst{The geometric transitions from the evolution in the complex plane of time provide channels for particle production for a quantum field in expanding universes. The production rate for one pair is obtained by squaring and summing the scattering matrix between the in-vacuum and the transported one over all possible independent closed paths of winding number 1.}

\kword{particle production, expanding universe, geometric transition, Stokes phenomenon}

\begin{document}
\maketitle

\section{Introduction} \label{sec 1}

Hawking radiation in a black hole, Gibbons-Hawking radiation in a de Sitter space, and Schwinger mechanism in a constant electric field are the most prominent particle production from the vacuum. The out-vacuum may be superposed of multi-particle states of the in-vacuum and their amplitudes square is the probability for those particles to be spontaneously created from the vacuum \cite{DeWitt75}. The phase-integral method, one of tunneling pictures, provides the probability for scattering over a barrier and penetration under a well, which give a physical intuition behind pair production \cite{Kim-Page07,Dumlu-Dunne10}.

Recently it has been proposed that particle production may be understood by extending the quantum evolution to the complex plane of time in the in-in formalism \cite{Kim13}. The idea is that the geometric contributions from the poles of an analytic Hamiltonian in the complex plane are responsible for not only particle production but also the Stokes phenomenon in de Sitter radiation \cite{Kim10}. In this paper we elaborate further and apply the geometric interpretation of particle production to expanding universes.

\section{Geometric Transitions and  Stokes Phenomenon}  \label{sec 2}

Let us consider the quantum evolution of a time-dependent oscillator
\begin{eqnarray}
\hat{H} (t) = \frac{1}{2M(t)} \hat{p}^2 + \frac{M(t) \omega^2 (t)}{2} \hat{q}^2,
\end{eqnarray}
with time-dependent frequency and/or mass, whose initial state propagates by the evolution operator expressed in the time-ordered integral (units of $\hbar = c =1$)
\begin{eqnarray}
\vert \psi (t) \rangle = {\rm T} \exp \Bigl[ - i \int_{t_0}^{t} \hat{H} (t') dt' \Bigr] \vert \psi (t_0) \rangle. \label{ev op}
\end{eqnarray}
On the real-time axis, the number states $ \hat{a}^{\dagger} (t) \hat{a} (t) \vert n, t \rangle = n \vert n, t \rangle$ are non-degenerate, provided that $\omega (t) > 0$, and constitute a basis. Then, arranging the energy spectrum as $\textbf{H}_{\rm D} (t) =  \omega (t) \text{diag} (1/2, \cdots, n+1/2, \cdots)$ and the energy eigenstates as $\Phi^{\rm T} (t) = (\vert 0, t \rangle, \cdots, \vert n, t \rangle, \cdots)$,
we may write the evolution operator in the form \cite{Kim13}
\begin{eqnarray}
\hat{U} (t, t_0) = \Phi^{\rm T} (t) {\rm T} \exp \Bigl[ - i \int_{t_0}^{t} \Bigl( \textbf{H}_{\rm D} (t') - \textbf{A}^{\rm T} (t')   \Bigr) dt' \Bigr]
\Phi^{*} (t_0),
\end{eqnarray}
where the induced gauge potential is an off-diagonal matrix $ \textbf{A} = i \Phi^* (\partial \Phi^T/\partial t)$.
Here ${\rm T}$ and $*$ denote the transpose and the dual operations and the time-ordered integral is understood as a matrix-valued operation,
which is not commutative in general. Hence the in-in scattering matrix in the real-time dynamics is trivial since $\langle \psi (t_0) \vert \psi (t_0) \rangle = 1$, unless there is a level crossing.

\begin{figure}[tbh]
\includegraphics[width=0.7\linewidth]{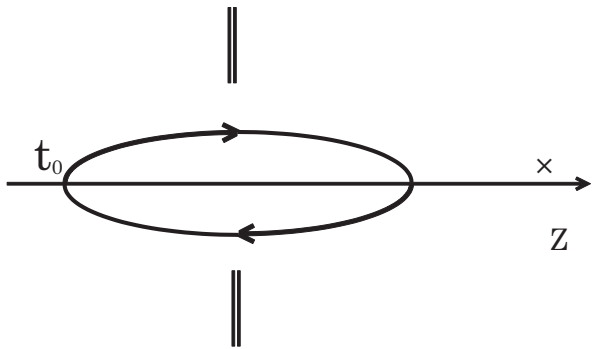}
\caption{The double lines denote two branch cuts.
The closed clockwise path in the complex plane $Z$ starts from the base point $t_0$ and either encloses a set of simple poles or excludes all simple poles. There is another simple pole at infinity marked by $\times$, which contributes a residue to any closed path of non-zero winding number.}
\label{fig-1}
\end{figure}

However, the rich structure in the in-in formalism can be revealed by extending the Hamiltonian $\hat{H} (z)$ to the whole complex plane of time.
Though requiring a rigorous mathematical analysis, we simply \textit{assume that an analytical operator $\hat{H}(z)$ exists in the whole complex plane,
reducing to the Hamiltonian $\hat{H}(t)$ along the real-time axis, and has an orthonormal basis $\langle n, z \vert m, z \rangle =\delta_{nm}$}.
To illustrate the essential concept of complex evolution, we consider a complex frequency
\begin{eqnarray}
\omega (z) =  f(z) \Bigl((z-z_+^*)(z-z_+)(z- z_-^*)(z-z_-) \Bigr)^{1/2}, \label{mod freq}
\end{eqnarray}
where $f(z)$ is an analytic function with a finite number of simple poles at $z_j$ and another simple pole at $z = \infty$. The pole at $z = \infty$ under a conformal mapping $z = z(t)$ covering the whole complex plane originates from external backgrounds and leads to the instanton action for particle production.
Two pairs of branch points for the square root occur in quantum motion of a massive particle in the global coordinates of a de Sitter space
and also of a charged particle in the Sauter-type electric field. The model frequency (\ref{mod freq}) exhibits an interesting feature, the so-called Stokes phenomenon, but may be extended to more general cases. To make (\ref{mod freq}) analytic in the complex plane, we cut the branch lines as shown by double lines in Fig. \ref{fig-1} that connect each pair. The analytic continuation in Fig. \ref{fig-1} contrasts the quantum evolution in this paper with that starting from a past real time to a future real time, whose path encloses branch points as level-crossings for geometric transitions \cite{Hwang-Pechukas77,JKP91}.

The scattering matrix between the in-vacuum at $t_0$ on the real-time axis and the transported one along a closed clockwise path $C(z)$ with the same base point $t_0$ in the complex plane is to the lowest order of the Magnus expansion
\begin{eqnarray}
\langle 0, t_0 \vert 0, C (t_0) \rangle = \exp \Bigl[- \frac{i}{2} \oint_{C (t_0)} \omega(z)dz \Bigr]. \label{sc am}
\end{eqnarray}
Then the production rate for one pair is the sum of the scattering matrix square over all independent paths \cite{Kim13}
\begin{eqnarray}
{\cal N} = \Bigl\vert \sum_{J} \langle 0, t_0 \vert 0, C^{(1)}_{J} (t_0) \rangle^2 \Bigr\vert =  \Bigl\vert \sum_{J} \exp \Bigl[-i \oint_{C^{(1)}_{J} (t_0)} \omega(z)dz \Bigr] \Bigl\vert. \label{pa pr}
\end{eqnarray}
Here $C^{(1)}_{J}$ exhausts all possible independent closed paths which are classified by the homotopy class for the simple poles and have the winding number 1. The higher winding number corresponds to multiple pair production.

To be more specific, denote the residues of simple poles by $\text{Res} (\omega (z_j))$ and $\text{Res} (\omega (\infty))$, respectively.  Then the particle-production rate takes the form
\begin{eqnarray}
{\cal N} =   e^{- 2 \pi \text{Res} (\omega (\infty))} \Bigl\vert \sum_{J} \exp \Bigl[- 2 \pi \sum_{j} \text{Res} (\omega (z_j)) \Bigr] \Bigl\vert. \label{pa for}
\end{eqnarray}
Note that the geometric contribution from the pole at infinity is universal and also that the particle-production rate involves double summation: a sum over independent paths and another sum of the residues from the poles inside the path.
There is always a closed path not enclosing any finite simple pole, whose geometric contribution is unity due to zero residue except for the contribution from the infinity. The real residues from finite simple poles either suppress particle production for positive signs or enhance particle production for negative signs. The pure imaginary residues at finite simple poles can result in constructive or destructive interference depending on the set of simple poles and their residues, which explains a geometric origin for the Stokes phenomenon for particle production \cite{Kim13}.

\section{Particle Production in Expanding Universe}  \label{sec 3}

A complex massive scalar $\phi$  in a (1+1)-dimensional Friedmann-Robertson-Walker universe with the conformal metric
\begin{eqnarray}
ds^2 = a^2 (\eta) (- d \eta^2 + dx^2), \label{con met}
\end{eqnarray}
has the Fourier-decomposed Hamiltonian
\begin{eqnarray}
H(\eta) = \sum_{\kappa} \frac{1}{2} \Bigl[\pi_{k}^*\pi_{k}+ \omega_{k}^2 (\eta) \phi^*_{\kappa} \phi_{\kappa}  \Bigr], \quad \omega_{\kappa}^2 (\eta) = k^2 + m^2 a^2 (\eta), \label{osc}
\end{eqnarray}
where $\pi_{k} = \dot{\phi}^*_{k}$ and $\pi^*_{k} = \dot{\phi}_{k}$.
To illustrate the geometric transitions for particle production, we consider two conformal spacetimes: (i) $a^2 (\eta) = (A + B \tanh (\rho \eta))^2$ and (ii) $a^2 (\eta) = A + B {\rm sech}^2 (\rho \eta)$, $(A \geq 0, B \geq 0)$. We map a Riemann sheet of complex time $\eta$ to the whole complex plane $z$ via the conformal mapping $e^{\rho \eta} = z, (- \pi/\rho < {\rm arg}\, \eta \leq \pi/ \rho)$.

In the case (i), the contour integral takes the form
\begin{eqnarray}
\oint \omega_k (\eta) d \eta = \frac{1}{\rho} \oint \sqrt{k^2 (z^2+1)^2 + m^2 \Bigl(A (z^2+1) + B (z^2 -1) \Bigr)^2 } \frac{dz}{z (z^2+1)}.
\end{eqnarray}
The simple pole at $z = 0$ from infinity and the finite simple pole at $z = i$ contribute to the contour integral while another pole at $z= -i$ is excluded for the causality reason, which yield
\begin{eqnarray}
{\cal N}_k = e^{- 2 \pi \omega_{k {\rm in}}/\rho } \Bigl(1 + e^{2 \pi m|B|/\rho} \Bigr), \quad \Bigl( \omega_{k {\rm in}} = \sqrt{k^2 + m^2 (A-B)^2}, \, \omega_{k {\rm out}} = \sqrt{k^2 + m^2 (A+B)^2} \Bigr). \label{pp m2}
\end{eqnarray}
The first term comes from a path not enclosing any of $\pm i$ and the second term from a path enclosing $i$. The result (\ref{pp m2}) is the same as the pair-production rate (47) of \cite{Kim-Page07} in a vector potential $A_{\parallel} (t) = - E_0 \tanh (t/T)$ and may be compared with (3.11) of \cite{Bernard-Duncan77} in the conformal metric $a^2 (\eta) = A + B \tanh (\rho \eta)$, $(A \geq B)$. In the second case (ii), the contour integral becomes
\begin{eqnarray}
\oint \omega_k (\eta) d \eta = \frac{1}{\rho} \oint \sqrt{(k^2 + Am^2) (z^2+1)^2 + 4 B m^2 z^2} \frac{dz}{z (z^2+1)}.
\end{eqnarray}
The simple poles are still located at $z = \pm i$, but their residues are pure imaginary. So the particle-production rate for $B > 0$ is
\begin{eqnarray}
{\cal N}_k = e^{- 2 \pi \omega_{k {\rm in}}/\rho } \Bigl\vert 1 + 2 e^{2i \pi m\sqrt{B}/\rho} + e^{4i \pi m\sqrt{B}/\rho} \Bigr\vert, \quad
\Bigl( \omega_{k {\rm in}} = \omega_{k {\rm out}}  = \sqrt{k^2 + m^2 A} \Bigr). \label{pp m3}
\end{eqnarray}
There is a constructive or destructive interference among four independent paths, which leads to the Stokes phenomenon.
In the opposite case of $B < 0$ for the scattering over a well, the particle-production rate has a similar form as (\ref{pp m2}). A model of particular interest is $B = \rho^2 p(p+1)$, which corresponds to the P\"{o}sh-Teller potential for a natural number $p$ and whose reflectionless scattering explains no particle production \cite{Kim11}. For large $p$ the particle-production rate is approximately given by
\begin{eqnarray}
{\cal N}_k = 4 \sin^2 (\pi p) e^{- 2 \pi \omega_{k {\rm in}}/\rho }, \label{pp m3-2}
\end{eqnarray}
and vanishes for a natural number $p$ as expected. Hence the Stokes phenomenon is a geometric effect of interference among independent paths in the complex plane, and similar phenomenon has been observed in Schwinger mechanism in certain electric fields \cite{Dumlu-Dunne10} and de Sitter radiation \cite{Kim10}.

\section{Conclusion}  \label{sec 4}

In the complex-time dynamics, the transported in-vacuum along a closed path of non-zero winding number carries useful information about particle production. The square of the scattering matrix between the in-vacuum and the transported one along a closed path of winding number 1 gives a channel for particle production for one pair, and the particle-production rate is the sum of all possible independent channels. Particle production and the Stokes phenomenon has a geometric interpretation.

\section*{Acknowledgment}
This paper was completed at Yukawa Institute for Theoretical Physics, Kyoto University. This work was supported in part by Basic Science Research Program through the National Research Foundation of Korea (NRF) funded by the Ministry of Education (NRF-2012R1A1B3002852).


\begin{thebibliography}{9}

\bibitem{DeWitt75} B.~S.~DeWitt: Phys.\ Rept.\ {\bf 19} (1975) 295.

\bibitem{Kim-Page07} S.~P.~Kim and D.~N.~Page: Phys.\ Rev.\ D \textbf{75} (2007) 045013.

\bibitem{Dumlu-Dunne10} C.~K.~Dumlu and G.~V.~Dunne: Phys.\ Rev.\ Lett.\ \textbf{104} (2010) 250402; C.~K.~Dumlu and G.~V.~Dunne: Phys.\ Rev.\ D \textbf{83} (2011) 065028; C.~K.~Dumlu and G.~V.~Dunne: Phys.\ Rev.\ D \textbf{84} (2011) 125023.

\bibitem{Kim13} S.~P.~Kim: Phys.\ Lett.\ B \textbf{725} (2013) 500; S.~P.~Kim: Phys.\ Rev.\ D \textbf{88} (2013) 044027.

\bibitem{Kim10} S.~P.~Kim: J.\ High Energy Phys.\ \textbf{09} (2010) 054.

\bibitem{Hwang-Pechukas77} J.-T.~Hwang and P.~Pechukas: J.\ Chem.\ Phys.\ \textbf{67} (1977) 4640.

\bibitem{JKP91} A.~Joye, H.~Kunz, and Ch.-Ed Pfister:  Ann.\ Phys.\ \textbf{208} (1991) 299.

\bibitem{Kim11} S.~P.~Kim: J.\ Korean\ Phys.\ Soc.\ \textbf{63} (2013), in press, arXiv:1110.4684.

\bibitem{Bernard-Duncan77} C.~W.~Bernard and A.~Duncan, Ann.\ Phys.\ \textbf{107} (1977) 201.

\end{thebibliography}
\end{document}